\documentclass[12pt,preprint]{aastex}
\usepackage{natbib}
\def\degree{\ifmmode {^\circ}\else {$^\circ$}\fi}
\def\rstar{\ifmmode {\, R_{\star}}\else $R_{\star}$\fi}
\def\msol{\ifmmode {\, M_{\odot}}\else $M_{\odot}$\fi}
\def\rsol{\ifmmode {\, R_{\odot}}\else $R_{\odot}$\fi}
\def\lsol{\ifmmode {\, L_{\odot}}\else $L_{\odot}$\fi}
\def\msolyr{\ifmmode {\, M_{\odot}\,{\rm yr}^{-1}}\else $M_{\odot}\,{\rm yr}^{-1}$\fi}
\def\mdot{\ifmmode {\,\dot{M}}\else $\dot{M}$\fi}
\def\mdotyr{\ifmmode {\,\dot{M}\,yr^{-1}}\else $\dot{M}\,yr^{-1}$\fi}

\newcommand{\Teff}{\ifmmode{T_{\rm eff}}\else{$T_{\rm eff}$}}

\begin{document}

\title{What are the R Coronae Borealis Stars?}

\author{ Geoffrey C. Clayton\altaffilmark{1}}

\altaffiltext{1}{Department of Physics \& Astronomy, Louisiana State
University, Baton Rouge, LA 70803; gclayton@fenway.phys.lsu.edu}

\begin{abstract}
The R Coronae Borealis (RCB) stars are rare hydrogen-deficient, carbon-rich, supergiants, best known for their spectacular declines in brightness at irregular intervals.
Efforts to discover more RCB stars have more than doubled the number known in the last few years and they appear to be members of an old, bulge population.  
Two evolutionary scenarios have been suggested for producing an RCB star, a double degenerate merger of two white dwarfs, or a final helium shell flash in a planetary nebula central star. The evidence pointing toward one or the other is somewhat contradictory, but the discovery that RCB stars have large amounts of $^{18}$O has tilted the scales towards the merger scenario. 
 If the RCB stars are the product of white dwarf mergers, this would be a very exciting result since
RCB stars would then be low-mass analogs of type Ia supernovae.
The predicted number of RCB stars in the Galaxy is consistent with the predicted number of He/CO WD mergers. But, so far, only about 65 of the predicted 5000 RCB stars in the Galaxy have been discovered. The mystery has yet to be solved.

\end{abstract}

% Keywords should be included, but they are not printed in the hardcopy.


\section{Introduction}
R Coronae Borealis (R~CrB) was one of the first variable stars identified and studied.
 It received the ``R" which designates it as the first variable star discovered in the constellation Coronae Borealis.
Its brightness variations have been monitored since its discovery over 200 years ago
\citep{1797RSPT...87..133P}. Early spectra
 showed variations in  the strengths of 
the Swan bands of $C_2$ \citep{1890MNRAS..51...11E} and evidence of the absence of hydrogen was soon detected \citep{1906AN....173....1L,1912AnHar..56...65C}, although not confirmed until later \citep{1935ApJ....81..369B,1953ApJ...117...25B}.
In addition, the explanation behind the large declines in brightness, the production of thick clouds of carbon dust, was deduced very early on \citep{1935AN....254..151L,1939ApJ....90..294O}.
But the stellar evolution that produced R~CrB remains mysterious.\\
\indent
The R Coronae Borealis (RCB) stars are a small group of carbon-rich supergiants. About 65
RCB stars are known in the Galaxy and 25 in the Magellanic Clouds.  
%\citep{1996PASP..108..225C,Alcock:2001lr,Zaniewski:2005ai,2005ApJ...631L.147K,Tisserand:2008kx}.
Their defining characteristics are hydrogen
deficiency and unusual variability. RCB stars undergo massive
declines of up to 8 mag due to the formation of carbon dust at
irregular intervals.  
The RCB stars can be roughly divided into a majority group which share similar abundances, and a small minority of stars, which show extreme abundance ratios, particularly Si/Fe and S/Fe \citep{Asplund:2000qy}.
There are also six hydrogen-deficient carbon (HdC) stars that are RCB stars
spectroscopically, but do not show declines in brightness or IR excesses
\citep{Warner:1967lr,2010ApJ...723L.238G,2012A&A...539A..51T}.\\
\indent
Two scenarios have been proposed for the origin of an RCB star: the double degenerate
(DD) and the final helium-shell flash (FF) models \citep{Iben:1996fj,2002MNRAS.333..121S}. The former involves
the merger of a CO- and a He-white dwarf (WD)
\citep{1984ApJ...277..355W}. In the latter case, thought to occur in 20\%
of all AGB stars, a star evolving into a planetary nebula (PN) central star undergoes a helium flash and
expands to supergiant size \citep{Fujimoto:1977lr}.
 Three stars
(Sakurai's Object, V605 Aql, and FG Sge) have been observed to
undergo FF outbursts that transformed them from hot evolved PN
central stars into cool giants with spectroscopic properties similar to
RCB stars \citep{1997AJ....114.2679C,1998ApJS..114..133G,1998A&A...332..651A,Asplund:1999bh,Asplund:2000qy,2006ApJ...646L..69C}.\\
\indent
Recent  observations and population synthesis models imply that there are a
significant number of close DD binaries in the Galaxy. A majority of
binaries, close enough to interact sometime during their evolution,
will end up as DD systems where both stars are WDs
\citep{2005A&A...440.1087N,2012ApJ...749L..11B}.
%The evolution of such a close binary system will result in two mass transfer phases, including at least one common envelope phase when the first star to become a WD is engulfed by the other star.
 If the resulting DD system has a short
enough period ($\lesssim$ 0.2 hr) it will enter a phase of
mass-transfer and may merge in less than a Hubble time due to the
loss of energy due to gravitational radiation. This may result in a
SN Ia explosion if the total mass of the DD system is greater than
the Chandrasekhar limit, or in RCB and HdC  stars if the mass is lower
than this limit \citep{1984ApJ...277..355W,1994ApJ...420..336Y}.
\\
\indent
Recently, the surprising
discovery was made that RCB stars have $^{16}$O/$^{18}$O ratios that are
orders of magnitude lower than for any other known stars \citep{2005ApJ...623L.141C,Clayton:2007ve,Garcia-Hernandez:2010fk}.
Greatly enhanced $^{18}$O is evident in every HdC and RCB that has been
measured  and is cool enough to have detectable CO bands. 
%In addition, \citet{Pandey:2008eu} have found that the abundance of $^{19}$F is 
%extremely high in the RCB stars. 
IR spectra of Sakurai's Object, obtained when it had
strong CO overtone bands, showed no evidence for $^{18}$O
\citep{2002Ap&SS.279...39G}. Therefore, Sakurai's Object and the
other FF stars on the one hand, and most of the RCB and HdC stars on
the other hand, are likely to be stars with different origins. 
No overproduction of $^{18}$O is expected in
the FF scenario but in a DD merger, partial helium burning may take place leading to enhanced  $^{18}$O \citep{Clayton:2007ve}.
This strongly suggests that the RCB stars are the results of mergers of close WD binary systems. 
These discoveries are important clues which will help to 
distinguish between the proposed DD and FF evolutionary pathways
of HdC and RCB stars. 
%KPD 1930+2752, a DD system consisting of a subluminous B star (sdB) and a WD has been suggested as a SN Ia progenitor candidate \citep{2007A&A...464..299G,2001A&A...376L...9E}. Recently, \citet{Izzard:2007ul} suggested that another type of carbon star --- the R-stars --- results from the merger of a WD and a red giant. Longer lived interacting DD systems, such as AM CVn binaries, will produce gravitational radiation that should be detected by the planned Laser Interferometer Space Antenna (LISA) mission \citep{2004MNRAS.349..181N}.
\\
\indent
There have been three conferences devoted to hydrogen-deficient stars held in 1985, 1995 and 2007. The proceedings of these conferences contain many papers concerning the RCB stars and related objects
\citep{1986ASSL..128.....H,1996ASPC...96.....J,2008ASPC..391.....W}.
The last general review of the RCB stars appeared 16 years ago \citep{1996PASP..108..225C}. Since then, the number of RCB stars known has more than doubled and about 250 papers have been written. This review will concentrate on the advances in our knowledge of RCB stars that have been made since 1996.

%\section{Population of RCB Stars}

\section{How to Identify an RCB star}

\noindent
A. \underline{The Lightcurve}

As seen in Figure 1, the RCB lightcurve is unique. No other type of star displays such wild, irregular, large amplitude variations \citep[e.g.,][]{1938HarMo...5.....P,1996PASP..108..225C}. \\
$\bullet$An RCB  star will stay at uniform brightness at maximum  for months or years. Then, there will be a sudden drop in brightness of more than three magnitudes taking a few days or weeks, followed by a recovery to maximum light, which is typically slower, taking months or years. Notice the latest decline of R~CrB, itself, shown in Figure 1. After over 1000 days at maximum, it plunged seven magnitudes in less than 100 days and has stayed in a deep decline for almost 2000 days. \\
$\bullet$The characteristic time between
declines in RCB stars is typically about 1000 days, but there is a wide range in activity
among the RCB stars \citep{1986ASSL..128..151F,1996AcA....46..325J}.\\
$\bullet$The stars also show regular or semi-regular pulsations with a small amplitude ($\Delta$V $\lesssim$0.1 mag) and periods of 40-100 days \citep[e.g.,][]{Lawson:1990fk,2004JAVSO..33...27P}.

\vspace{0.5cm}
\noindent
B. \underline{The Spectrum}

Figure 2 shows the maximum light spectra of the RCB stars, S Aps (T$_{eff}\sim$5000 K) and RY Sgr (T$_{eff}\sim$7000 K), and  the carbon star, RV Sct.\\
$\bullet$ The RCB spectra are characterized by weak or absent hydrogen Balmer lines, and weak or absent CH $\lambda$4300 band. But, there is a wide range of hydrogen abundance in the RCB stars. For example, V854 Cen shows significant hydrogen in its spectrum \citep{1989MNRAS.238P...1K,Lawson:1989kx}. There is an anti-correlation between hydrogen and metallicity in the RCB stars \citep{Asplund:2000qy}.\\
$\bullet$ The RCB star spectrum contains many lines of neutral 
atomic carbon. The cooler stars (T$_{eff}<$6000 K) show strong absorption bands of C$_2$ and CN, as seen in S Aps, but in the warmer stars, such as RY Sgr, these bands are weak or absent. A warm RCB star can appear almost featureless in the visible, having no Balmer lines, helium lines, or molecular bands.\\
$\bullet$ In most, but not all, RCB stars, the abundance of $^{13}$C is very small. This can be seen in the two wavelength sections shown in Figure 2. The isotopic Swan bands of  C$_2$ are separated in wavelength, so the spectra can be inspected to see the relative strengths of $^{12}$C$^{12}C$ $\lambda$4737.0 and $^{12}$C$^{13}$C $\lambda$4744.0 in the blue, and $^{12}$C$^{12}$C $\lambda$$\lambda$6059, 6122, 6191, and $^{12}$C$^{13}$C $\lambda$$\lambda$6100, 6168 in the red \citep{Lloyd-Evans:1991qy,Kilkenny:1992fk,Alcock:2001lr}.\\
$\bullet$ The strength of the CN $\lambda$6206 band compared to the $^{12}$C$^{12}$C $\lambda$6191 band is relatively weak in the RCB stars compared to the carbon stars \citep{Lloyd-Evans:1991qy,Morgan:2003yq}.
 \\
\indent
The description above applies to RCB spectra at or near maximum light. In deep declines,
a rich 
narrow-line emission spectrum appears consisting of lines of neutral and singly 
ionized metals, and a few broad lines including Ca II H and K, 
the Na I D lines  \citep{1963ApJ...138..320P,1972MNRAS.158..305A,1975IAUS...67..129F}. The decline spectrum is described in detail in \citet{1996PASP..108..225C}. More recent papers detailing the decline spectra of RCB stars include,
\citet{Rao:1997,Rao:1999,1999ApJ...515..351C,1999MNRAS.302..341S,1999AJ....117.3007L,2000MNRAS.313L..33K,2001ApJ...560..986C,2001IBVS.5063....1K,2002MNRAS.335.1133S,2002Obs...122..322S,2004MNRAS.355..855K,2006BaltA..15..521K,Kameswara-Rao:2006lr}.
 A small subclass of RCB stars is much hotter with effective temperatures of about 20,000 K. See 
\citet[][and references therein]{2002AJ....123.3387D} for a description of their spectra.

\vspace{0.5cm}
\noindent
C. \underline{Infrared Emission}\\
\noindent
$\bullet$Every RCB star has an IR excess, from the K-band to far-IR wavelengths due to warm circumstellar dust ($T_{eff}$ $\sim$ 400 - 1000 K) \citep{Feast:1997lr}. Most Galactic RCB stars have been detected by 2MASS, IRAS, AKARI, and WISE \citep[e.g.,][]{Walker:1985rr,2012A&A...539A..51T}.
Recently, two RCB stars, R~CrB and HV 2671, were detected out to 500 \micron~by the Herschel Space Observatory \citep{Clayton:2011lr,2011ApJ...743...44C}. \\
$\bullet$The mid-IR spectra of RCB stars are mostly featureless since there are usually no silicate, SiC, or polycyclic aromatic hydrocarbon (PAH) features present 
\citep{2001ApJ...555..925L,2005ApJ...631L.147K,2011ApJ...739...37A}. However, V854 Cen and DY Cen do show emission features attributed to PAHs and C$_{60}$ \citep{2001ApJ...555..925L,2011ApJ...729..126G}. 

\section{The Population of RCB Stars}

\citet{1996PASP..108..225C} listed 34 Galactic and 3 LMC RCB stars. Since then, the number of confirmed RCB stars has almost doubled in the Galaxy, and greatly increased  to over 25 in the Magellanic Clouds. 
Tables 1 and 2 list all the RCB and HdC stars known in the Galaxy and the Magellanic Clouds which have been confirmed by spectral classification, lightcurve behavior, and IR excesses.
V2331 Sgr is a strong new candidate from its lightcurve and IR excess, but does not have a spectrum (Tisserand et al. 2012 in preparation). V581 CrA, EROS2-LMC-RCB-6 and OGLE-GC-RCB-2 are also good candidates that do not have spectra \citep{Tisserand:2009fj,Tisserand:2011uq,2012A&A...539A..51T}.
V391 Sct was originally classified as a dwarf nova that brightened from V=17 to 13 mag. But Brian Skiff (personal communication) suggested that this star might be an RCB star based on brightness variations seen on a few plates. This is supported by the ASAS-RCB-3 lightcurve which, while it does not show any declines, reveals that the star is normally V=13 not V=17. Its spectrum shows it to be a warm RCB star, very similar to RY Sgr and it has an IR excess  (Tisserand et al. 2012 in preparation).  A strong spectroscopic candidate, KDM 6546, has no lightcurve \citep{Morgan:2003yq}. It was previously classified as a CH star \citep{1988ApJ...334..135H}. Three stars included in the RCB list of \citet{1996PASP..108..225C}, GM Ser, V1773 Oph, and V1405 Cyg, had not been spectroscopically confirmed \citep{1997Obs...117..205K}. GM Ser is not an RCB star (Tisserand et al. 2012 in preparation). The other two stars are still unconfirmed and so are not included in Table 1. Another star, MACHO 118.18666.100, previously identified as an RCB star \citep{Zaniewski:2005ai}, has been shown to be misidentified \citep{Tisserand:2008kx}.
 \\
\indent
There are also four hot (15,000--20,000 K) RCB stars known. One, HV 2671, was recently discovered in the LMC \citep{1996ApJ...470..583A}. The three Galactic stars are, V348 Sgr, MV Sgr and DY Cen. These stars are all hydrogen-deficient, carbon-rich stars, and have RCB-type lightcurves \citep{1996PASP..108..225C,2002AJ....123.3387D,Clayton:2011lr}.  DY Cen and MV Sgr have the typical RCB-star large helium abundances, but V348 Sgr and HV 2671 are in general agreement with a born-again post-AGB evolution, and are similar to Wolf-Rayet central stars of PNe with carbon and helium being close to equal in abundance  \citep{2002AJ....123.3387D}.
So DY Cen and MV Sgr seem to be related to the cooler RCB stars, but V348 Sgr and HV 2671 may be [WC] central stars.
The six known HdC stars are also listed in Table 1. One of these stars, HD 175893, may be an RCB star since it has an IR excess \citep{2012A&A...539A..51T}. However, no declines have been observed for this star. \\
\indent
There is a small group of stars, of which DY Per is the prototype, that resemble the RCB stars \citep{1994BaltA...3..410A}. These stars have very deep declines at irregular intervals, but the rate of fading is very slow and the shape of the declines is much more symmetrical than the typical RCB decline \citep{Alcock:2001lr}. DY Per is significantly cooler (T$_{eff}\sim$3500 K) than the coolest RCB stars \citep{1997PASP..109..969K}. 
\\
\indent
 Both of the
evolutionary theories, the DD and the FF scenarios, suggest that the
RCB stars are an old population \citep{1996PASP..108..225C}. 
The distribution on the sky and radial velocities of the RCB stars tend toward those of the bulge population \citep{1986ASSL..128....9D,1998PASA...15..179C,Zaniewski:2005ai}.
\citet{Tisserand:2008kx} determined that the RCB stars follow a disk-like distribution inside the Bulge with a scale-height $<$250 pc. 
The distribution of the RCB stars on the sky, including the new expanded sample from Table 1, is plotted in Figure 3.
There is no direct measurement of the distance to any Galactic RCB star \citep{1996ApJ...470..583A}.
But, now that significant numbers of RCB stars have been identified in the LMC, whose distance is well known, the absolute brightness of the  RCB stars has been determined to range from M$_V$= -3 for stars with T$_{eff}\sim$5000 K to -5 for stars with T$_{eff}\sim$7000 K \citep{1979ctvs.conf..246F,Alcock:2001lr,Tisserand:2009fj}. HV 2671, the hot RCB star in the LMC, has M$_V\sim$-3 \citep{Alcock:2001lr,Tisserand:2009fj}. 
\\
\indent
\citet{1984ApJ...277..355W} suggested that the DD
scenario would result in a population of about 1000 Galactic RCB stars.
 \citet{Iben:1996fj} put the
Galactic RCB population resulting from the same scenario at about
300 stars, and calculated that the FF scenario would imply anywhere
from 30 to 2000 RCB stars at any given time.  All of the evidence
thus far suggests that there are many more than the $\sim$65 known RCB
stars in the Galaxy \citep[e.g.,][]{Zaniewski:2005ai}.
The number of RCB stars expected in the Galaxy can be extrapolated from the LMC RCB population, using the method described by
\citet{Alcock:2001lr}, and including all the new LMC stars. This implies a population of RCB stars in the Galaxy of almost 5,700.
RCB stars are thought to be $\sim$0.8-0.9  M$_{\sun}$ from pulsation modeling \citep{Saio:2008qe}, and this mass agrees well with the predicted mass of the merger products of a CO- and a He-WD \citep{Han:1998kl}. On the other hand, FF stars, since they are single WDs, should typically have masses of 0.55-0.6 M$_{\sun}$ \citep{Bergeron:2007tg}.
\\
\indent
%The estimated production of RCB stars from mergers \citep{Han:1998kl} is 0.018 yr$^{-1}$.
The merger rate of He+CO DD's is predicted to be
$\sim 0.018 ~\mathrm{yr}^{-1}$ in the Galaxy \citep{Han:1998kl}. As of 1988, only one such DD  system was
actually known to exist \citep{1988ApJ...334..947S}. The SPY project and other surveys
have studied many WDs for evidence of binarity and the number of
known DD systems is now $\sim$100 
\citep[e.g.,][]{2005A&A...440.1087N,2010ApJ...716..122K,2011MNRAS.413L.101K,2011ApJ...735L..30P,2011ApJ...737L..23B,2012ApJ...749L..11B}.
To see how well this number matches the predicted number of RCB stars in the Galaxy, we need to estimate how long an RCB star formed from a DD merger will live.
This lifetime can be calculated as, t = $\Delta M \times X \times Q/L$,
where L is the luminosity of
RCB stars, $\Delta$M is the accreted mass of He, X is the mass fraction of He
in the accreted material and Q is the energy generated when one gram of
He is burned to $^{12}$C.
Assuming that Q$\sim7 \times10^{18}$ erg, $\Delta$M = 0.1M$_{\sun}$, X=0.3, and L= 10,000 L$_{\odot}$, t$\sim3\times 10^5$ yr. Using the estimate of  \citet{Han:1998kl} for the production of RCB stars from DD mergers, then the predicted population of RCB stars in the Galaxy at any given time would be $\sim$5,400 which agrees well with the number extrapolated from the LMC above.

\section{Evolutionary History}

Table 3 summarizes the observational data that must be addressed by evolutionary models of RCB stars \citep{Asplund:2000qy,2011MNRAS.414.3599J}. They are discussed below with respect to the FF and DD models. Two entries, the high abundance of silicon and sulphur, and the anti-correlation of hydrogen with iron cannot be well explained by either scenario. The condensation of certain elements into dust has been suggested for the Si/S problem, although it is unclear that this would work \citep{Asplund:2000qy}. The H/Fe anti-correlation is unexplained but Sakurai's Object follows the same trend.

\subsection{Do RCB Stars Evolve from Final Flash stars?}

The lightcurve behavior and spectroscopic appearance of V605 Aql in 1921 and more recently of Sakurai's Object are reminiscent of the RCB class. There are, however, several reasons why
FF stars are unlikely to be the evolutionary precursors of the majority of RCB stars.
The FF objects have deeper light declines ($>$10~mag) than do RCB stars ($\sim$8~mag). 
This may be due to the fact that RCB stars have more dust lying near the star which scatters light around the intervening dust cloud. This may account for the flat-bottom appearance of deep RCB declines. See Figure 1. 
The 
abundances of FF objects, shortly after the outburst, do appear similar to those of RCB stars, except for some interesting differences.
First there are significant amounts of $^{13}C$ present in Sakurai's Object, but not in most RCB stars. 
In the two years after it appeared, Sakurai's Object had $^{12}C/^{13}C \sim 4$ \citep{1994MNRAS.268..544P,Asplund:1999bh,2004A&A...417L..39P}.
In general, RCB stars have $^{12}C/^{13}C \geq 100$.
However, a few RCB stars do have detectable  $^{13}C$ including both majority and minority stars. V CrA, V854 Cen, VZ Sgr, and UX Ant have measured $^{12}$C/$^{13}$C $<$25 \citep{2008MNRAS.384..477R,2012ApJ...747..102H}. Second, there is no sign of $^{18}$O in the IR CO bands of Sakurai's Object \citep{1998MNRAS.298L..37E}.  
Finally, several Galactic and LMC RCB stars, including R~CrB, itself, show significant Lithium in their atmospheres \citep{Rao:1996oq,Asplund:2000qy,Kipper:2006fk}. 
\citet{Renzini:1990wd} suggested that in a FF the ingestion of the H-rich envelope leads to Li-production through the Cameron-Fowler mechanism \citep{Cameron:1971lr}.
The abundance of Li in the atmosphere of the FF star, Sakurai's object, was actually observed to increase with time \citep{Asplund:1999bh}.
\\
\indent
In general, Sakurai's abundances resemble V854 Cen and the other minority-class RCB stars with 98\% He and 1\% C
  \citep{1998A&A...332..651A}. Although, only low resolution spectra are available for V605 Aql from 1921, it likely had similar abundances \citep{1997AJ....114.2679C}.
  New spectra obtained of V605 Aql in 2000 indicate that it has evolved from T$_{eff}\sim$5000 K in 1921 to $\sim$95,000 K today \citep{2006ApJ...646L..69C}. The new spectra also show that V605 Aql now has stellar abundances similar to those seen in [WC] central stars of PNe with $\sim$55\% He, and $\sim$40\% C. 
In the present state of evolution of V605 Aql, we may be seeing the not too distant future of Sakurai's Object. 
There are indications that Sakurai's Object is evolving along a similar path \citep{2002ApJ...581L..39K,2005Sci...308..231H}.
Some doubt has recently been thrown on the FF nature of V605 Aql on the grounds of high neon abundances found in its ejecta  \citep{2008MNRAS.383.1639W,2011MNRAS.410.1870L}.
\\
\indent
For a very short time, perhaps as short as two years, both V605 Aql and Sakurai's Object were almost indistinguishable from the RCB stars.
Unfortunately, this extremely short RCB phase of the FF stars means that they cannot account for even the small number of RCB stars known in the Galaxy. 
%From R~CrB itself, we have a lower limit on the lifetime of an RCB star of $>$200 yr \citep{1996PASP..108..225C}. 

\subsection{Do RCB Stars Evolve from White Dwarf Mergers?}

RCB and HdC stars have $^{16}$O/$^{18}$O ratios close to
and in some cases less than unity, values that are orders of magnitude
lower than measured in any other stars (the Solar value is 500) \citep{2005ApJ...623L.141C,Clayton:2007ve,Garcia-Hernandez:2010fk}. The three HdC stars, that have been measured, have $^{16}$O/$^{18}$O~$<$~1, lower values than any of the RCB stars. These discoveries are important clues in determining the evolutionary
pathways of HdC and RCB stars, whether the DD or the
FF. No overproduction of $^{18}$O is expected in the FF scenario. 
%We also find qualitative agreement with the high observed values of $^{12}$C/$^{13}$C and with the CNO elemental ratios. 
%H-admixture during the accretion process from the small H-rich C/O WD envelope may play an important role in producing the observed abundances. 
New hydrodynamic simulations indicate that WD mergers may very well
be the progenitors of O$^{18}$-rich RCB and HdC stars \citep{2011ApJ...737L..34L,staff12}.
\\
\indent
\citet{1984ApJ...277..355W} proposed that an RCB star evolves from the
merger of a He-WD and a CO-WD which has passed through a common
envelope phase. He suggested that as the binary begins to coalesce
because of the loss of angular momentum by gravitational wave
radiation, the (lower mass) He-WD is disrupted. A fraction of the
helium is accreted onto the CO-WD and starts to burn, while the
remainder forms an extended envelope around the CO-WD. This structure,
a star with a helium-burning shell surrounded by a 
$\sim$100~R$_\odot$ hydrogen-deficient envelope, closely resembles an RCB star \citep{1996PASP..108..225C,Clayton:2007ve}.
The merging times ($\sim$10$^9$ yr) might not be as long as previously thought, which makes
the DD scenario an appealing alternative to the FF scenario for the formation of RCB stars
\citep{Iben:1996fj,1997ApJ...475..291I}.  In addition, \citet{2002MNRAS.333..121S} suggested that a
WD-WD merger could also account for the elemental abundances seen in RCB
stars.
\citet{2006ApJ...638..454P} have suggested a similar origin for the EHe stars. 
\\
\indent
The isotope, $^{18}$O, can be overproduced in an
environment of partial He-burning in which the temperature and the
duration of nucleosynthesis are such that the
$^{14}N$($\alpha,\gamma$)$^{18}$F($\beta^+$)$^{18}$O reaction chain can produce
$^{18}$O, if the $^{18}$O is not further processed by
$^{18}$O($\alpha,\gamma$)$^{22}$Ne \citep{Warner:1967lr,1986hdsr.proc..127L,Clayton:2007ve}.
The surface compositions of HdC and RCB stars are extremely He-rich (mass
fraction 0.98), indicating that the surface material has been processed
through H-burning. After H-burning via the CNO cycle, $^{14}$N is by far the
most abundant of the CNO elements, because $^{14}$N has the smallest nuclear
p-capture cross-section of any stable CNO isotope involved. However, the majority of RCB stars has 
$\log{C/N} = 0.3$
and $\log{N/O} = 0.4$ by number, equivalent to mass ratios of $C/N =
1.7$ and $N/O = 2.2$ \citep{Asplund:2000qy}. The N/O ratio represents the average for the majority RCB stars although the individual stars show a large scatter. Thus C is the most
abundant and O the least abundant CNO element. These abundances are
consistent if the material at the surfaces of
HdC and RCB stars experienced a small amount of He-burning, as, for
example, at the onset of a He-burning event that is quickly
terminated. This partial He-burning would not significantly deplete He,
but could be sufficient for some of the $^{14}$N to be processed into $^{18}$O.
%Other abundance signatures of RCB and HdC stars are a high \czw/\cdr\
%ratio ($>500$) and s-process enhancement \citep{asplund:00}. Li
%overabundances have also be observed in some objects of this class.  
At the onset of He-burning, $^{13}$C is the first $\alpha$-capture
reaction to be activated because of the large cross-section of $^{13}$C($\alpha$,n)$^{16}$O.  Thus, a large $^{12}$C/$^{13}$C ratio and
enhanced s-process elements are both consistent with partial He-burning.
\\
\indent
As mentioned above, some RCB stars show enhanced Li
abundances, as does the FF star Sakurai's Object \citep{1986hdsr.proc..127L,1998A&A...332..651A}. As shown by \citet{2001NuPhA.688..221H}, Li enhancements are consistent with the FF
scenario. However, the production of $^{18}$O requires temperatures
large enough to at least marginally activate the
$^{14}$N$(\alpha,\gamma)$ reaction. The $\alpha$ capture on Li is eight
orders of magnitude more effective than on $^{14}$N. For that reason, the
simultaneous enrichment of Li and $^{18}$O is not expected in the WD merger scenario. This is an
important argument against the FF evolution scenario as a progenitor
of RCB and HdC stars with excess of $^{18}$O. 
The abundance of $^{18}$O cannot be directly measured in R~CrB because it is too hot to have CO, but it  is overabundant in $^{19}$F, which does imply a high $^{18}$O abundance \citep{Pandey:2008eu}.
Since
$^{18}$O strongly supports the DD merger/accretion scenario, 
the obvious solution is that there could be (at least) two 
evolutionary channels leading to RCB, HdC and EHe stars, perhaps
with the DD being the dominant mechanism. Unfortunately the
division between majority- and minority-class RCB stars does not lend itself naturally to this explanation,
since Li  has only been detected in
the majority group \citep{Asplund:2000qy}.

\subsection{Mass-Loss and Dust Formation}

It has long been accepted that characteristic RCB declines in brightness are caused by the formation of optically thick clouds of carbon dust \citep{1935AN....254..151L,1939ApJ....90..294O,1996PASP..108..225C}. But the formation mechanism is not well understood. Empirical analysis of the spectroscopic and lightcurve evolution during declines implies that the dust forms close to the stellar atmosphere and then is accelerated to hundreds of km s$^{-1}$~by radiation pressure \citep{1992ApJ...397..652C,1993ASPC...45..115W}.
There is strong evidence for variable, high-velocity winds in RCB stars associated with dust formation
\citep{2003ApJ...595..412C,clayton12a}.
The HdC stars, which produce no dust, also have no evidence for winds \citep{2009ApJ...698..735G}.
Other observational evidence indicates that there is also gas moving much more slowly away from the star \citep[e.g.,][and references therein]{2011ApJ...739...37A}.
\\
\indent
The observed timescales for 
RCB dust formation fit in well with those calculated by carbon chemistry models which show that the dust forms near the surface of the RCB star due to density and temperature perturbations caused by stellar pulsations \citep{1986ASSL..128..151F,1996A&A...313..217W}. 
There is a strong correlation between the onset of dust formation and the pulsation phase in several RCB stars \citep{2007MNRAS.375..301C}.
All RCB stars show regular or semi-regular pulsation periods in the 40-100 d range \citep{Lawson:1990fk}.
The dust forming around an RCB star does not form in a complete shell, but rather in small ``puffs" \citep{1996PASP..108..225C}. Only when a puff forms along the line of sight to the star will there be a decline in brightness. Studies of UV extinction and IR re-emission of stellar radiation, indicate that the covering factor of the clouds around RCB stars during declines is f $<$ 0.5 \citep{Feast:1997lr,1999ApJ...517L.143C,1984ApJ...280..228H,1998ApJ...501..813H}. 
The typical dust mass of a puff is $\sim$10$^{-8}$ M$_{\sun}$ \citep{1986ASSL..128..151F,1992ApJ...397..652C}.
\\
\indent
 In the recent deep decline of R~CrB, shown in Figure 1, the puff dust mass is $\sim$10$^{-8}$ M$_{\sun}$, 
assuming the dust forms at 2 R$_{\star}$ (R$_{\star}$=85 R$_{\sun}$), and that a puff subtends a fractional solid angle of 0.05 \citep{2011ApJ...743...44C}. Since the dust would accelerate and dissipate quickly due to radiation pressure, dust must be formed continually by R~CrB to maintain itself in a deep decline for 4 years or more. If a puff forms during each pulsation period ($\sim$40 d), R~CrB would be producing $\sim$10$^{-7}$ M$_{\sun}$ of dust per year. Assuming a gas-to-dust ratio of 100 \citep{2003dge..conf.....W}, the total mass loss rate is 10$^{-5}$ M$_{\sun}~yr^{-1}$.
\\
\indent
Little is known about the lifetime of an
RCB star.  We have a lower limit from the fact that R~CrB itself has been an RCB star for 200 yr \citep{1797RSPT...87..133P}.
The large diffuse dust shell around R~CrB, seen in the far-IR,  could possibly be a fossil PN shell  \citep{2011ApJ...743...44C}. 
Assuming an expansion velocity of a PN shell ($\sim$20 km s$^{-1}$), then the shell would take  $10^5$ yr to form. If the mass-loss was more like the high-velocity winds seen in RCB stars today ($\sim$200 km s$^{-1}$), then the shell would be about an order of magnitude younger \citep{2003ApJ...595..412C}. 
If R~CrB is the result of a FF rather than a DD, then the size and timescales would be consistent with the nebulosity, now seen in far-IR emission, being a fossil PN shell. The nebulosity including cometary knots, seen around R~CrB and UW Cen looks very much like a PN shell \citep{1999ApJ...517L.143C,2011ApJ...743...44C}. The mass of the shell is estimated to be $\sim$2 M$_{\sun}$, which is consistent with it being a PN \citep{2011ApJ...743...44C}.
Adaptive optics and interferometry have been used to study dust very close to RCB stars \citep[][and references therein]{de-Laverny:2004lr,2011MNRAS.414.1195B}.
\\
\indent
 Any gas lost during a DD event would have far less mass. 
 If the shell is an old PN shell then this would suggest that R~CrB is the product of an FF event rather than a DD merger.
R CrB is one of the stars with lithium on its surface, also a possible indicator of an FF.
 %The lifetime of R~CrB as an RCB star cannot be much longer than this because it is a low-mass star. 
 %There exist a few RCB stars that show significant Lithium in their spectra, which may instead favor the FF model \citep{1996ApJ...456..750I,1996PASP..108..225C}. 
 About 10\% of single stars will undergo a final-flash event \citep{Iben:1996fj}.
%The {\it IRAS} data for 16 RCB stars have been re-examined \citep{walker94}. 
About this percentage of RCB stars
(R~CrB, RY Sgr, V CrA, and UW Cen) show evidence of resolved fossil dust shells  \citep{walker94}. %This fits with the prediction of \citet{Iben:1996fj}.
\\
\indent
Understanding the RCB and HdC stars is a key test for any theory that aims
to explain hydrogen deficiency in post-AGB stars.
Solving the mystery of how the RCB stars evolve is an exciting possibility, but it will also be a watershed event in the study of stellar evolution that could lead to a better understanding of other types of stellar merger events such as type Ia  supernovae.

The observations in the AAVSO database have been invaluable to my research on RCB stars throughout my career. For R CrB alone, there are a staggering 238,136 brightness estimates in the database, stretching back to 1843. The usefulness of the AAVSO data is only increasing with the addition of digital data which allow the low-amplitude RCB star pulsations to be studied in detail. The AAVSO database is a model for the many other photometric databases coming on line. The need for longterm monitoring combined with reliable photometry is essential for the identification and characterization of the many transient objects that are being discovered.

\acknowledgments
We acknowledge with thanks the variable star observations from the AAVSO International Database contributed by observers worldwide and used in this research. I would especially like to thank Albert Jones for his many visual estimates of the RCB stars. This work has been supported, in part, by grant  NNX10AC72G from NASAÕs Astrophysics Theory Program.

\bibliography{/Users/gclayton/projects/latexstuff/everything2}
\clearpage

\begin{deluxetable}{lcccccccc}
%\small
%\rotate
\tabletypesize{\scriptsize}
\tablewidth{0pt}
\tablecaption{Spectroscopically Confirmed Milky Way RCB stars }
\tablehead{\colhead{Name}
&
\colhead{$\alpha(2000)$} &
\colhead{$\delta(2000)$} 
& \colhead{Max}&
\colhead{SpT Ref$^1$}& 
\colhead{$^{18}$O}&\colhead{F}&\colhead{Li}&
\colhead{Notes$^2$}}
\startdata
%DY Per&SR&02 35 05&+56 09 12&12.6&1&&&&\nodata\\
XX Cam&04 08 38.75&  +53 21 39.5&8.7&1&&x&x&\nodata\\
SU Tau&05 49 03.73 & +19 04 22.0&9.5&1&&\checkmark&\checkmark&\nodata\\
UX Ant&10 57 09.06 & -37 23 55.0&12.2&1&&x&x&$^{13}C$\\
UW Cen&12 43 17.18  &-54 31 40.7&9.6&1&x&\checkmark&\checkmark&\nodata\\
Y Mus&13 05 48.19 & -65 30 46.7&10.5&1&x&&x&\nodata\\
DY Cen&13 25 34.08  &-54 14 43.1&12.0&1&&&&hot RCB\\
V854 Cen&14 34 49.41 & -39 33 19.2&7.0&1&x&x&x&Minority, $^{13}C$\\
Z UMi&15 02 01.33 & +83 03 48.6&11.0&1&x&&\checkmark&Minority\\
S Aps&15 09 24.53 & -72 03 45.1&9.6&1&\checkmark&&&\nodata\\
ASAS-RCB-1  &   15 44 25.08& -50 45 01.2&11.9&2&&&&V409 Nor\\
R~CrB&15 48 34.41 & +28 09 24.3&5.8&1&&\checkmark&\checkmark&\nodata\\
ASAS-RCB-9 &  16 22 28.83 &-48 35 55.8 &11.3&2,3&&&&IO Nor\\
RT Nor&16 24 18.68 & -59 20 38.6&11.3&1&&&x&\nodata\\
RZ Nor&16 32 41.66 & -53 15 33.2&11.1&1&&&\checkmark&\\
ASAS-RCB-2  &   16 41 24.75 &-51 47 43.4 &12.0&2&&&&\nodata\\
ASAS-RCB-3  &   16 54 43.60 &-49 25 45.0 &11.8&2,3&&&&\nodata\\
ASAS-RCB-12&17 01 01.41& -50 15 34.8&11.8&2&&&&\nodata\\
ASAS-RCB-4   &  17 05 41.25 &-26 50 03.4 &13.3&2,3&&&&GV Oph\\
V517 Oph&17 15 19.74 & -29 05 37.6&12.6&1&&&&\nodata\\
ASAS-RCB-10   & 17 17 10.21& -20 43 15.7&11.5&2&&&&\nodata\\
%Vl773 Oph&17 17 22&--20 22 36&16.8&1&&&&\nodata\\
EROS2-CG-RCB-12&17 19 58.50&-30 04 21.3&14.1&4&&&&\nodata\\
V2552 Oph&17 23 14.55 & -22 52 06.3&10.8&5,6&&\checkmark&x&\nodata\\
EROS2-CG-RCB-7&17 29 37.09&-30 39 36.7&14.1&4&&&&\nodata\\
EROS2-CG-RCB-6 & 17 30 23.83&-30 08 28.3&12.8&4&&&&V1135 Sco\\
V532 Oph&17 32 42.61&-21 51 40.8&11.7&7&&&&\nodata\\
OGLE-GC-RCB-1 &  17 35 18.12& -26 53 49.2&14.6&8&&&&\nodata\\
EROS2-CG-RCB-8&17 39 20.72&-27 57 22.4&13.0&4&&&&\nodata\\
EROS2-CG-RCB-10&17 45 31.41&-23 32 24.4&12.5&4&&&&\nodata\\
EROS2-CG-RCB-5&17 46 00.32&-33 47 56.6&13.5&4&&&&\nodata\\
EROS2-CG-RCB-4&17 46 16.20&-32 57 40.9&12.5&4&&&&\nodata\\
EROS2-CG-RCB-9&17 48 30.87&-24 22 56.5&15.2&4&&&&\nodata\\
EROS2-CG-RCB-11&17 48 41.53&-23 00 26.5&12.3&4&&&&\nodata\\
ASAS-RCB-7    & 17 49 15.70& -39 13 16.0 &12.7&2&&&&V653 Sco\\
EROS2-CG-RCB-1&17 52 19.96&-29 03 30.8&12.4&4&&&&\nodata\\
ASAS-RCB-5  &   17 52 25.30 &-34 11 28.0 &12.3&2&&&&\nodata\\
EROS2-CG-RCB-2&17 52 48.70&-28 45 18.9&14.5&4&&&&\nodata\\
MACHO 401.48170.2237&17 57 59.02&-28 18 13.1&14.5&9&&&&\nodata\\
EROS2-CG-RCB-3&17 58 28.27&-30 51 16.4&11.1&4&&&&\nodata\\
EROS2-CG-RCB-13&18 01 58.22&-27 36 48.3&11.4&4&&&&\nodata\\
V1783 Sgr&18 04 49.74 & -32 43 13.4&12.5&2&&&&\nodata\\
%GM Ser &18 08 36&--15 04 00&12.0&no sp&&&&\nodata\\
WX CrA&18 08 50.48 & -37 19 43.2&11.0&1&\checkmark&&&\nodata\\
ASAS-RCB-11  &  18 12 03.30 &-28 08 33.0&12.0&2&&&&\nodata\\
V739 Sgr&18 13 10.54 & -30 16 14.7&14.0&1&&&&\nodata\\
EROS2-CG-RCB-14&18 13 14.86&-27 49 40.9&12.5&4&&&&\nodata\\
V3795 Sgr &18 13 23.58 & -25 46 40.8&11.5&1&&\checkmark&&Minority\\
VZ Sgr&18 15 08.58 & -29 42 29.4&11.8&1&&\checkmark&x&Minority, $^{13}C$\\
IRAS 18135-2419&18 16 39.20 &-24 18 33.4&12.8&2,10&&&&\nodata\\
RS Tel&18 18 51.22 & -46 32 53.4&9.3&1&&&x&\nodata\\
MACHO 308.38099.66&18 19 27.36&-21 24 08.2&16.3&9&&&&\nodata\\
MACHO 135.27132.51&18 19 33.87&-28 35 57.8&14.3&9&&&&\nodata\\
GU Sgr &18 24 15.58 & -24 15 26.5&11.3&1&&\checkmark?&x&\nodata\\
V581 CrA&18 24 43.46&	-45 24 43.8&10.0&2&&&&\nodata\\
V391 Sct&18 28 06.57 &-15 54 44.7&13.3&2&&&&\nodata\\
MACHO 301.45783.9&18 32 18.60&-13 10 48.9&17.3&9&&&&\nodata\\
NSV 11154&18 37 51.26& +47 23 23.5&12.0&11&&&&\nodata\\
V348 Sgr&18 40 19.93 & -22 54 29.3&10.6&1&&&&hot RCB\\
MV Sgr&18 44 31.97 & -20 57 12.8&12.0&1&&&&hot RCB\\
FH Sct &18 45 14.84 & -09 25 36.1&13.4&1&&\checkmark?&x&\nodata\\
V CrA&18 47 32.30 & -38 09 32.3&9.4&1&&\checkmark?&x&$^{13}$C\\
ASAS-RCB-8  &  19 06 39.87&-16 23 59.2&10.9&2&&&&\nodata\\
SV Sge&19 08 11.76  &+17 37 41.2&11.5&1&\checkmark&&&\nodata\\
V1157 Sgr&19 10 11.83&  -20 29 42.1&12.5&1&&&&\nodata\\
RY Sgr&19 16 32.76 & -33 31 20.4&6.5&1&&\checkmark&x&\nodata\\
ES Aql&19 32 21.61  &-00 11 31.0&11.6&12&\checkmark&&&\nodata\\
V482 Cyg&19 59 42.57&  +33 59 27.9&12.1&1&&\checkmark?&x&\nodata\\
ASAS-RCB-6   & 20 30 04.96&	-62 07 59.2 &13.1&2,3&&&&AN 141.1932\\
%V1405 Cyg&21 57 31&+53 53 42&15.5&no sp&&&&????\\
U Aqr &22 03 19.70 & -16 37 35.2&10.5&1 &\checkmark&&\checkmark?&\nodata\\
UV Cas&23 02 14.62 & +59 36 36.6&11.8&1&&\checkmark&x&\nodata\\
\sidehead{{\bf HdC Stars}}
HE 1015-2050&10 17 34.232&-21 05 13.87&16.0&13&&&&\nodata\\
HD 137613&15 27 48.316& -25 10 10.15&7.5&14&\checkmark&&&\nodata\\
HD 148839&16 35 45.788& -67 07 36.69&8.3&14&x&&\checkmark&\nodata\\
HD 173409&18 46 26.627& -31 20 32.07&9.5&14&x&&&\nodata\\
HD 175893&18 58 47.29& -29 30 18.08&9.4&14&\checkmark&&&IR Excess\\
HD 182040&19 23 10.08& -10 42 11.54&7.0&14&\checkmark&&&\nodata\\
\enddata
\tablenotetext {1}{Spectroscopic References: 
1) \citet[][and references therein]{1996PASP..108..225C}, 
2) Tisserand et al. (2012 in preparation)
3) \citet{2012arXiv1204.4181M}, 
4) \citet{Tisserand:2008kx},
5) \citet{2003PASP..115.1301H}, 
6) \citet{Rao:2003vn}, 
7) \citet{2009PASP..121..461C}, 
8) \citet{Tisserand:2011uq},
9) \citet{Zaniewski:2005ai}, 
10) \citet{2007PZ.....27....7G},
11) \citet{2011PASP..123.1149K},
12) \citet{2002PASP..114..846C}, 
13) \citet{2010ApJ...723L.238G}, 
14) \citet{Warner:1967lr}
}
\tablenotetext {2}{Notes to Table 1: RZ Nor has a faint blue star nearby. This may explain why the RZ Nor declines are not very deep and it appears bluer during declines. }
%$\bullet$ V1773 Oph, GM Ser, V1405 Cyg: There is no published 
%spectroscopic confirmation of these stars.}
\end {deluxetable}

\begin{deluxetable}{lllcllllc}
\tabletypesize{\scriptsize}
\tablewidth{0pt}
\scriptsize
\tablecaption{Spectroscopically Confirmed LMC and SMC RCB stars }
\tablehead{\colhead{Name} 
&
\colhead{$\alpha(2000)$} &
\colhead{$\delta(2000)$} 
& \colhead{Max}&
\colhead{Spec. Ref.$^1$}& 
\colhead{$^{18}$O}&\colhead{F}&\colhead{Li}&
\colhead{Notes}}
\startdata
\sidehead{{\bf LMC Stars}}
EROS2-LMC-RCB-3&04 59 35.78&-68 24 44.68&14.3&1&&&&\nodata\\
HV 12524&05 01 00.36 &-69 03 43.2&14.5&2&&&&MACHO 18.3325.148\\
KDM 2373 &05 10 28.50&-69 47 04.3&13.8&1,3&&&&MACHO 5.4887.14,EROS2-LMC-RCB-2\\
HV 5637&05 11 31.37 &-67 55 50.6&15.8&4&&&&MACHO 20.5036.12\\
EROS2-LMC-RCB-1 &05 14 40.17&-69 58 40.1&15.2&1&&&&MACHO 5.5489.623\\
HV 2379 &05 14 46.20& -67 55 47.4&14.9&2&&&&MACHO 16.5641.22\\
MACHO 79.5743.15&05 15 51.79 &-69 10 08.6&15.2&2&&&&\nodata\\
 MACHO 6.6575.13&05 20 48.21& -70 12 12.5&15.3&2&&&&\nodata\\
 HV 942 &05 21 48.00 &-70 09 57.4&15.0&2&&&&MACHO 6.6696.60\\
 MACHO 80.6956.207&05 22 57.37 &-68 58 18.9&16.0&2&&&&\nodata\\
W Men &05 26 24.52 &-71 11 11.8&13.8&4&&&\checkmark&MACHO 21.7407.7\\
MACHO 80.7559.28&05 26 33.91 &-69 07 33.4&15.8&2&&&&\nodata\\
MACHO 81.8394.1358&05 32 13.36& -69 55 57.8&16.3&5&&&&\nodata\\
HV 2671  &05 33 48.94& -70 13 23.4&16.1&5&&&&MACHO 11.8632.2507, Hot RCB\\
EROS2-LMC-RCB-4&05 39 36.97 &-71 55 46.4&15.1&1&&&&MACHO 27.9574.93\\
KDM 5651 &05 41 23.49&-70 58 01.8&14.4&3&&&&MACHO 15.9830.5\\
HV 12842&05 45 02.88& -64 24 22.7&13.7&4&&&\checkmark&\nodata\\
 MACHO 12.10803.56&05 46 47.74 &-70 38 13.5&15.1&2&&&&\nodata\\
KDM 7101&06 04 05.53&-72 51 23.1&14.2&1,3&&&&EROS2-LMC-RCB-5\\
%EROS2-LMC-RCB-6&06 12 10.48&-74 05 10.16&&4,5&&&&no sp\\
\sidehead{{\bf SMC Stars}}
RAW 21&00 37 47.40&-73 39 02.0&15.6&1,3,6&&&&EROS2-SMC-RCB-1\\
%MH95 431&00 40 14.65&-74 11 21.2&16.6&5&&&&\nodata\\
 %RAW 233&00 44 07.45&-72 44 16.3&16.6&5&&&&\nodata\\
 RAW 476&00 48 22.87&-73 41 04.7&15.5&1,6&&&&EROS2-SMC-RCB-2\\
 EROS2-SMC-RCB-3&00 57 18.12&-72 42 35.3&16.0&1,6&&&&MACHO 207.16426.1662\\
\enddata
\tablenotetext {1}{Spectroscopic References: 
1) \citet{Tisserand:2009fj},
2) \citet{Alcock:2001lr}, 
3) \citet{Morgan:2003yq},
4) \citet[][and references therein]{1996PASP..108..225C}, 
5) \citet{1996ApJ...470..583A}, 
6) \citet{2004A&A...424..245T}}
%\tablenotetext {2}{Notes to Table 1}
%$\bullet$ V1773 Oph, GM Ser, V1405 Cyg: There is no published 
%spectroscopic confirmation of these stars.}
\end {deluxetable}

\begin{deluxetable}{lll}
%\small
%\rotate
\tabletypesize{\scriptsize}
\tablewidth{0pt}
\tablecaption{DD vs FF$^1$}
\tablehead{\colhead{Property}&
\colhead{DD}
&
\colhead{FF}
}
\startdata
Extreme H deficiency but some H present&yes?&yes\\
H abundance anti-correlated with Fe&?&?\\
Li abundance high in 5 stars (all majority)&no&yes\\
C/He $\sim$ 1\%&yes&no\\
 $^{12}$C/$^{13}$C $>$ 500&yes&no\\
 High N, O&yes&yes\\
 High Na, Al&yes?&yes\\
 High Si, S&?&?\\
 Enrichment of s-process elements&yes?&yes\\
 Abundance uniformity/non-uniformity for majority/minority&no?/yes&yes/no?\\
 Similar to Sakurai's object&no&yes\\
 Nebulosity present in a few stars&yes?&yes\\
 RCB Lifetime&yes&no\\
 Lack of binarity&yes&no?\\
  $^{18}$O and $^{19}$F greatly enhanced in (all?) stars &yes&no\\
  M$_V$= -3 to -5 mag&yes&yes\\
  Mass = 0.8 - 0.9 M$_{\sun}$&yes&no?\\
\enddata
\tablenotetext {1}{Adapted and updated from Table 7 of  \citet{Asplund:2000qy}.}
%\tablenotetext {2}{\citep{1996PASP..108..225C,Asplund:2000qy,2006ApJ...646L..69C,Clayton:2007ve,Alcock:2001lr}}
%$\bullet$ V1773 Oph, GM Ser, V1405 Cyg: There is no published 
%spectroscopic confirmation of these stars.}
\end {deluxetable}

\clearpage

\begin{figure}
%\figurenum{1} 
\begin{center}
\includegraphics[width=5in,angle=0]{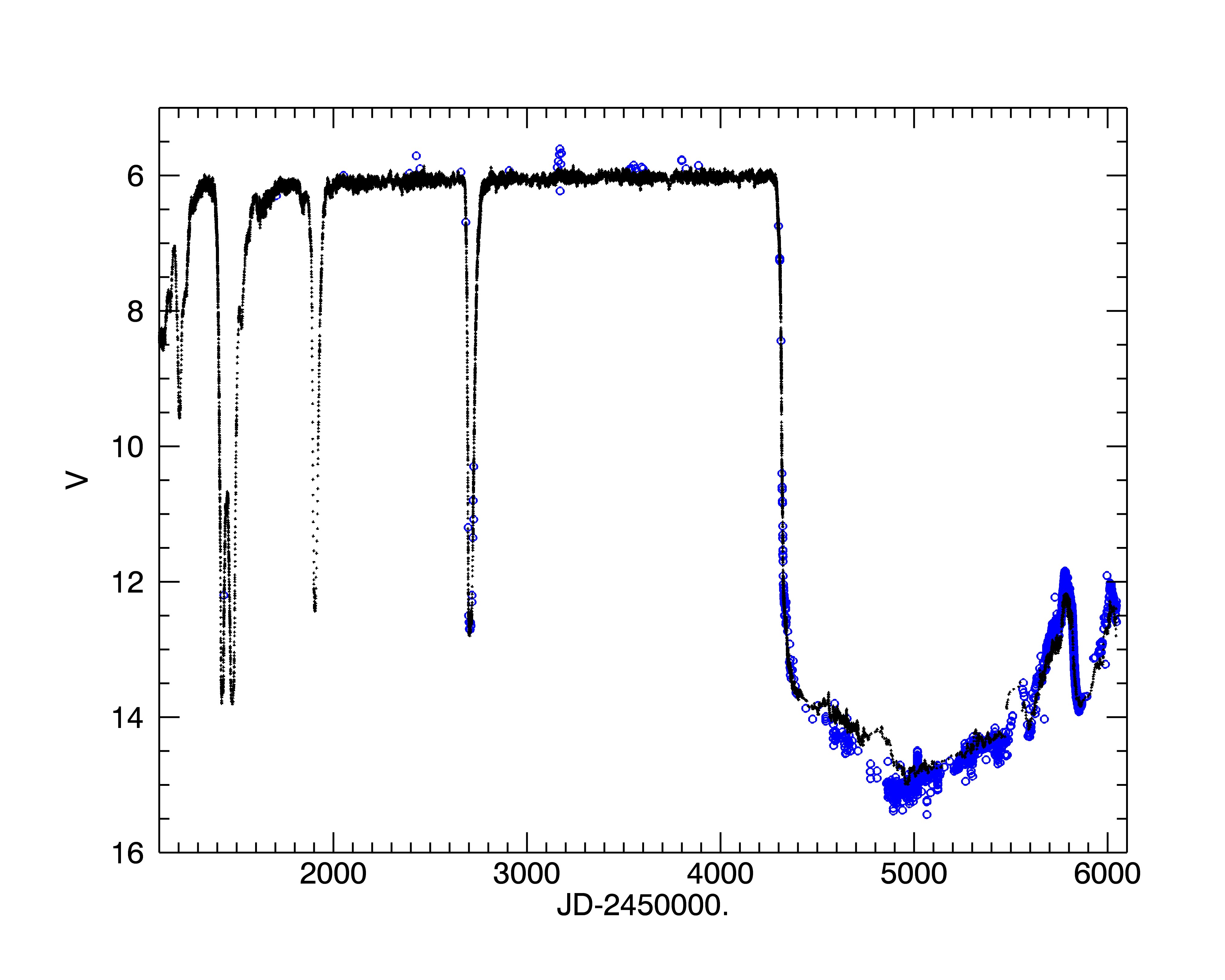}
\end{center}
\caption{Lightcurve of R~CrB from 1998-2012 using AAVSO data. Visual magnitudes are plotted as black dots. Johnson V data are plotted as blue open circles. \label{fig1}}
\end{figure}

\begin{figure}
%\figurenum{1} 
\begin{center}
\includegraphics[width=5in,angle=0]{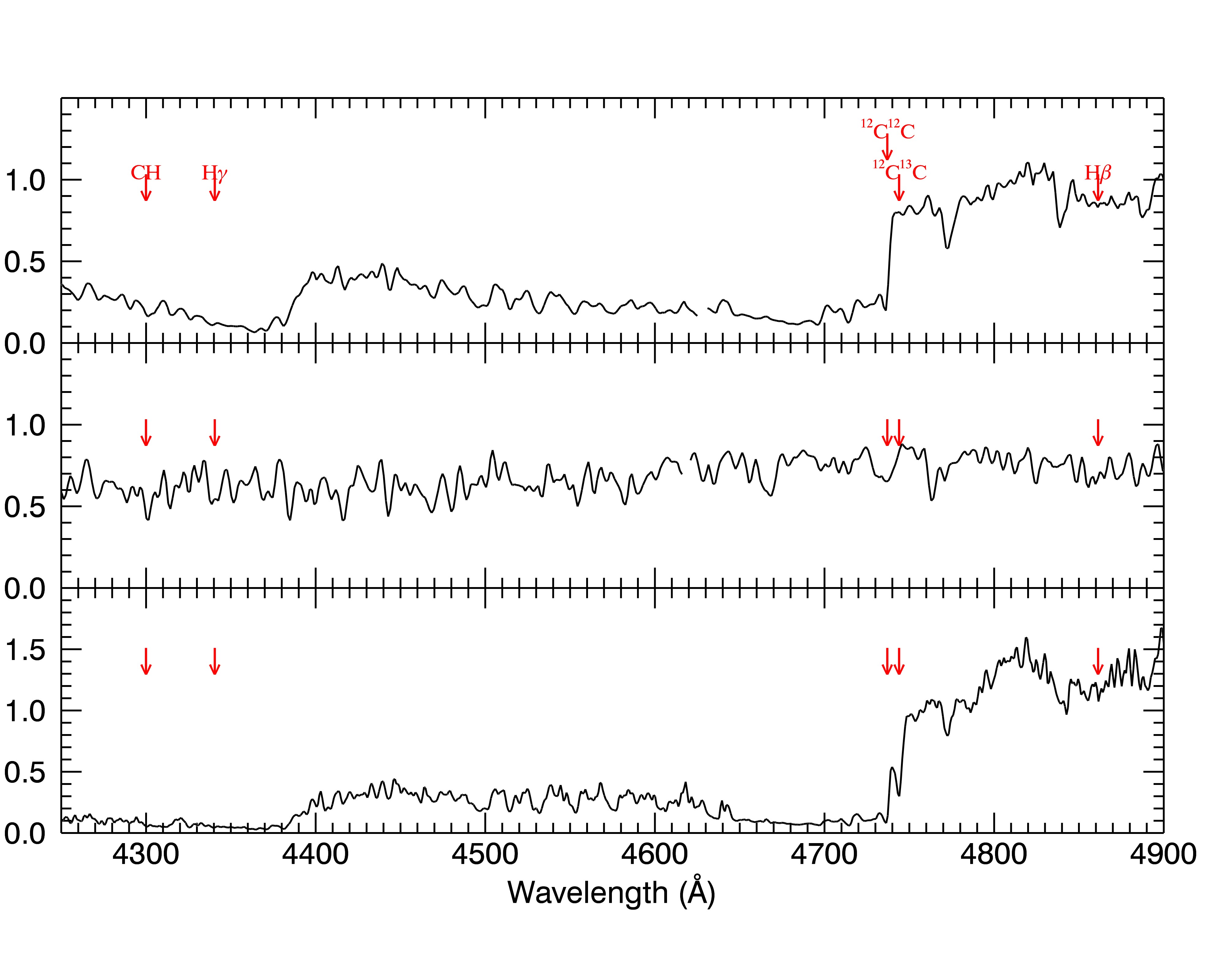}
\includegraphics[width=5in,angle=0]{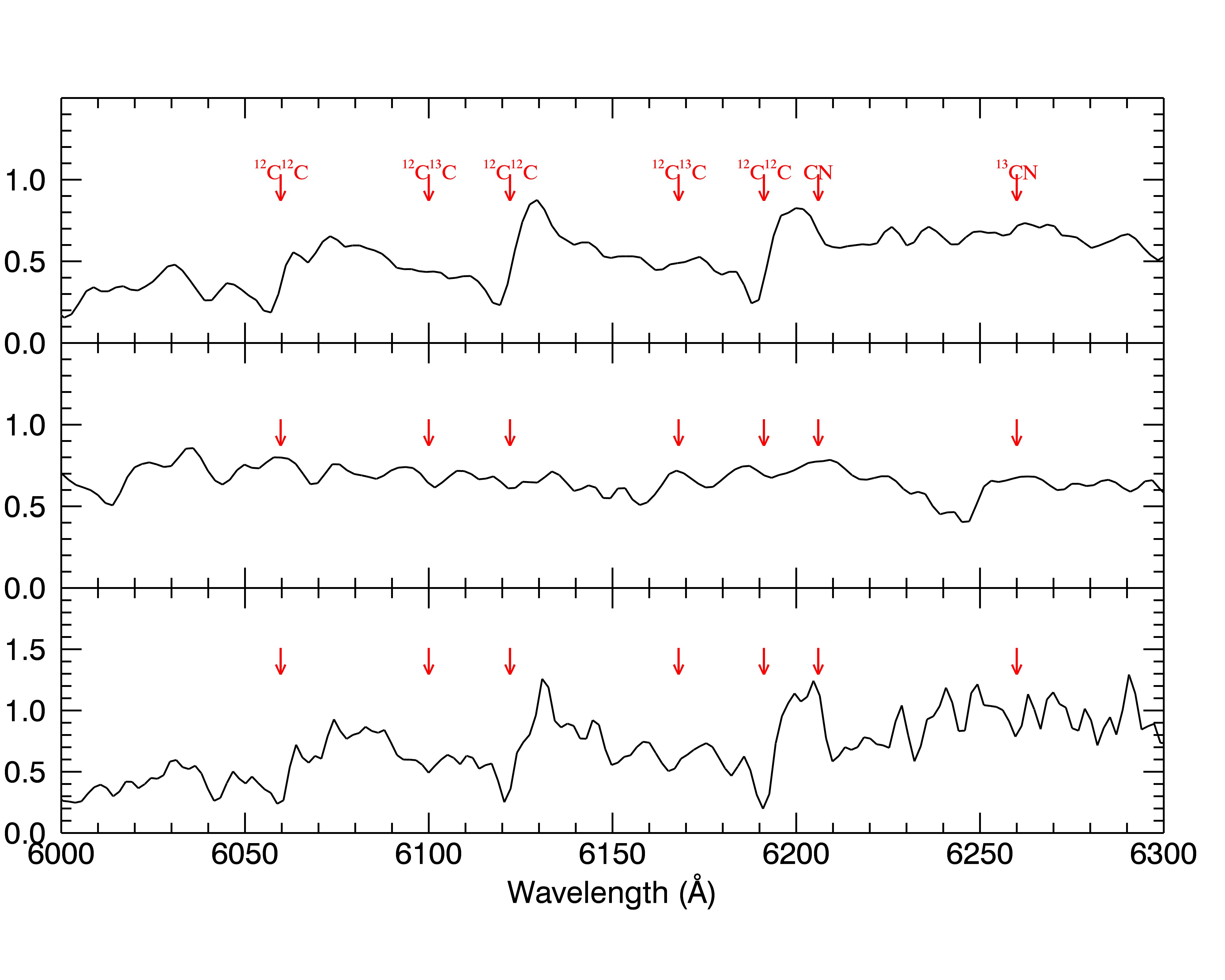}
\end{center}
\caption{Blue and red sections of the spectra of a cool RCB stars, S Aps (top), and a warm RCB star, RY Sgr (middle), as well as the carbon star, RV Sct (bottom), showing some of the spectroscopic features that define RCB stars.} 
\end{figure}

\begin{figure}
%\figurenum{1} 
\begin{center}
\includegraphics[width=5in,angle=0]{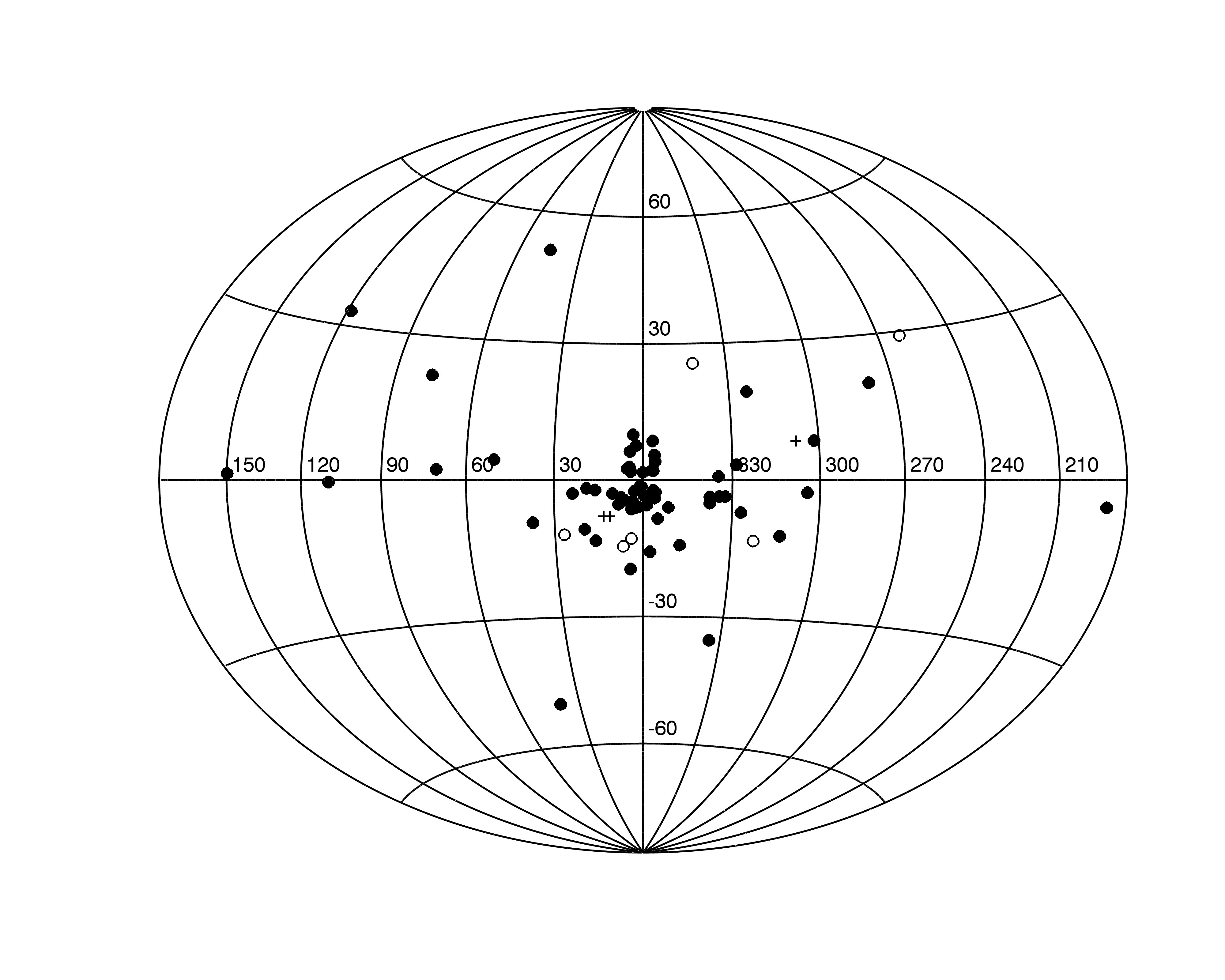}
\end{center}
\caption{Distribution of RCB stars on the sky in the Galaxy showing that they are consistent with an old, bulge population. Cool RCB stars (filled circles), hot RCB stars (crosses), HdC stars (open circles)}
\end{figure}

\end{document}